\documentclass[a4paper,prl,twocolumn]{revtex4}
\usepackage{graphicx}
\usepackage{subfigure}
\usepackage{amssymb}
\usepackage{color}
\begin{document}
\title{The effect of ion-induced shock waves on the transport \\ of reacting species around energetic ion tracks}

\author{Pablo de Vera$^{1,3,4}$}

\email[Corresponding author: ]{p.devera@qub.ac.uk}

\author{Eugene Surdutovich$^{2}$}

\author{Nigel J. Mason$^{3}$}

\author{Fred J. Currell$^{1}$}

\author{Andrey V. Solov'yov$^{4}$}

\email[On leave from A. F. Ioffe Physical Technical Institute, 194021 St. Petersburg, Russian Federation.]{}

\affiliation{$^{1}$ School of Mathematics and Physics, Queen's University Belfast, BT7 1NN Belfast, Northern Ireland, UK}

\affiliation{$^{2}$ Departament of Physics, Oakland University, Rochester, Michigan 48309, USA}

\affiliation{$^{3}$ Department of Physical Sciences, The Open University, MK7 6AA Milton Keynes, England, UK}

\affiliation{$^{4}$ MBN Research Center, Altenh\"{o}ferallee 3, 60438 Frankfurt am Main, Germany}

\begin{abstract}
The passage of energetic ions through tissue initiates a series of physico-chemical events, which lead to biodamage. The study of this scenario using a multiscale approach brought about the theoretical prediction of shock waves initiated by energy deposited within ion tracks. These waves are being explored in this letter in different aspects. The radial dose that sets their initial conditions is calculated using diffusion equations extended to include the effect of energetic $\delta$-electrons. The resulting shock waves are simulated by means of reactive classical molecular dynamics. The simulations predict a characteristic distribution of reactive species which may have a significant contribution to biodamage, and also suggests experimental means to detect the shock waves.

\end{abstract}

\maketitle

The basic understanding of the interaction of energetic ions with biomaterials is important both for radiotherapy and radiation protection from natural or human-made sources (cosmic radiation during manned space travel, Earth natural radioactivity or nuclear reactors)~\cite{Cucinotta2006}. In radiotherapy, proton and heavier ion beams have been exploited since the 1990s in the advanced technique referred to as ion-beam cancer therapy (IBCT)~\cite{Schardt2010,Loeffler2013,Surdutovich2014,Solovyov2017}. Macroscopically, ion beams feature a depth-dose curve where the maximum of energy loss (the Bragg peak) is reached close to the end of ions trajectories, allowing a precise energy delivery to the tumor region while sparing surrounding healthy tissues. Microscopic patterns of energy deposition around each ion path feature extremely high radial doses which steeply decrease on a nanometer scale. In the Bragg peak region, secondary electrons, free radicals, and other reactive species are produced in large numbers, leading to much higher (compared to photon radiation) concentrations of DNA lesions and formation of multiply-damages sites. These processes increase the probability of cell death or sterilization because of the suppressed capability of enzymes to repair such a complex damage.

A thorough understanding of radiation damage with ions is needed to develop a reliable optimization for IBCT treatment planning. The relation of the cell survival probability to the physical dose is non-trivial. In a common x-ray therapy a single physical parameter, the dose, is involved, but the biological diversity is so staggering that empirical components in all existing models is essential. A number of molecular models have been developed since the 1960s with an effort to mathematically explain the experimental dependence of cell survival probabilities on dose~\cite{Alpen}. The microdosimetric kinetic model (MKM)~\cite{MKM} is one of the most advanced approaches of that kind; it predicts cell survival as a function of dose and linear energy transfer (LET). A popular local effect model (LEM)~\cite{Scholz1996,Schardt2010} relates the radial dose to the cell survival using its relation to the dose for cells irradiated with x-rays. Another approach is being pursued by the track-structure community, with the idea of including all relevant processes, from ionization/excitation of the medium with ions to nuclear DNA damage, using Monte Carlo (MC) simulations~\cite{StewartMCDS,Friedland2017}.

Another alternative still is a multiscale approach (MSA)~\cite{Solovyov2009,Surdutovich2014,Solovyov2017} that took a radically different direction. Instead of starting with the observed cell survival probabilities, pertinent physical, chemical, and biological processes are analyzed combining a variety of temporal, spatial and energy scales. One important distinction from the track-structure approach is the prediction of shock waves initiated by each ion propagating in the medium. The strength of these shock waves
increases with LET, so they might be a substantial part of the radiation damage scenario around the Bragg peak region. Thus far, these shock waves have not been observed directly, but there is an evidence that makes their existence plausible. The detected acoustic waves from the Bragg peak region~\cite{Baily} are likely to be the artifacts of shock waves. Moreover, a successful comparison of experimental cell survival with the MSA, which included the shock waves in the scenario, for a variety of cell lines, values of LET, oxygen environments and cell repair conditions~\cite{Verkhovtsev2016}, is impressive.

The predicted shock waves are a consequence of the localized energy transfer from ions to the medium, due to the propagation of secondary electrons (most of them having very low energies of $\sim$45 eV in the vicinity of the Bragg peak \cite{deVera2013,deVera2013b} and thus travelling just a few nanometers) and the lack of mechanisms to quickly propagate this energy away from ions paths. Indeed, the diffusion mechanism is too slow and the production of energetic $\delta$-electrons in the Bragg peak vanishes~\cite{Surdutovich2014,Surdutovich2015}. Thus the development of high pressures inside of a narrow cylinder is expected. This happens by the time of $\sim10^{-13}$~s after the ion traverse, when all electrons have been thermalized, and suggests an onset of the cylindrical shock wave propagating radially away from the ion path. {Thus, the shock wave propagates on the time scale just between the so-called physical and chemical stages of radiation, typically considered to be fairly well separated in time, and both preceding the biological stage \cite{Mozumder2004}. While the evolution of track structure finishes by $\sim10^{-14}$~s (end of physical stage) and reactive chemical species (free radicals and solvated electrons) are created by $\sim10^{-13}$--$10^{-12}$~s, the chemical stage is deemed to start by $\sim10^{-12}$~s after irradiation and last until $\sim10^{-6}$~s. However, the emerging waves may have a substantial influence on the chemistry of the scenario, effectively mixing the physical and chemical stages of irradiation.}

Classical molecular dynamics (MD) simulations were previously used to investigate the direct mechanical effect of these waves on radiation damage \cite{Yakubovich2012,Surdutovich2013,deVera2016}. It was shown that covalent bonds can be ruptured by stress from the shock wave if the target is close enough to the ion path and the LET is sufficiently large. In Ref.~\cite{Surdutovich2013} it was also predicted that shock waves may play a significant role in the transport of reactive species such as free radicals due to radial collective flows initiated by them. Even though this idea has been heavily exploited in Refs.~\cite{Surdutovich2014,Verkhovtsev2016} it has never been properly studied. The analysis of formation and diffusion of free radicals in Ref.~\cite{Surdutovich2015} suggests that if there are no shock waves, most of the radicals do not leave ion tracks since they annihilate due to high rates of chemical reactions, high concentrations and the inability of the diffusion mechanism to steer them outside.



This letter reports results of MD simulations of transport of hydroxyl radicals by a shock wave. After the initial distributions of energy deposition and radicals around the ion path is obtained by solving the diffusion equations for the propagation of electrons, MD is used to simulate the shock waves. We investigate the effect of the radial-dose including $\delta$-electrons on the strength of the shock waves both in and out of the Bragg peak region.
Then the reactive force field implemented in MBN Explorer~\cite{Solovyov2012,Sushko2016} allows us to simulate one of the most representative chemical reactions occurring around the ion path (OH recombination) in the presence of the wave.

\begin{figure}[t]
\centering
\includegraphics[width=1.0\columnwidth]{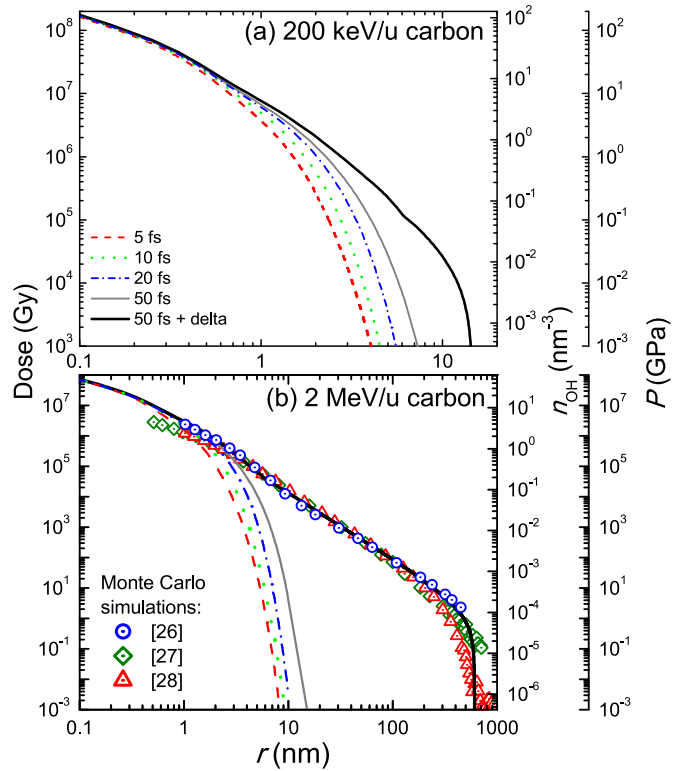}
\caption{(Color online) Left axis: radial doses produced in liquid water by (a) 200-keV/u and (b) 2-MeV/u carbon ions. Thin lines represent the time evolution of radial dose calculated using the diffusion equations, while thick lines represent final radial dose distributions including the $\delta$-electrons contribution. Symbols correspond to MC simulations results~\cite{Waligorski1986,Liamsuwan2013,Incerti2014}. Right axes: radial pressure and OH radical distributions produced by ions.
}
\label{fig:fig12}
\end{figure}

In Refs.\cite{Solovyov2009,Surdutovich2014,Surdutovich2015} the transport of electrons in the vicinity of the Bragg peak was treated analytically using the diffusion equations
and the radial dose deposition profile at a given time was calculated~\cite{Surdutovich2015}.
In Fig.~\ref{fig:fig12} the time evolution of the radial dose produced by a carbon ion (a) in the Bragg peak region (energy 200-keV/u) and (b) at 2-MeV/u (out of the Bragg peak) calculated by the diffusion equations is shown by thin lines. The radial dose increases with time until it saturates at $\sim$50 fs; then all electrons thermalize.
It should be noted that there is no mechanism by which the energy deposited so quickly can be dissipated gradually, since processes such as electron-phonon interaction or diffusion take place in much longer time scales~\cite{Gerchikov2000,Surdutovich2015}.

A limitation of the diffusion equations is that, treating all first generation electrons as having the average kinetic energy of $45$~eV, they cannot reproduce a large-radii dose tail coming from the contribution of balistic $\delta$-electrons, much less common but more energetic (see MC results for 2-MeV/u carbon ions in water in Fig. \ref{fig:fig12}(b)). In Ref.~\cite{Surdutovich2015} $\delta$-electrons were not accounted for explicitly, since their production is supressed in the Bragg peak region.
However, to also simulate the effect of shock waves out of the Bragg peak, in this work we have implemented a recipe to account for $\delta$-electrons based on a spatially-restricted LET formula~\cite{Xapsos1992}, while the low-energy electrons are still treated by the diffusion equations. The technical implementation goes beyond the goal of this letter and it will be described in detail in another work \cite{deVera2017}.

The radial doses at the end of the track-structure, including $\delta$-electrons, are represented in Fig.~\ref{fig:fig12} by thick lines; they are compared to different MC simulations for 2-MeV/u carbon ions \cite{Waligorski1986,Liamsuwan2013,Incerti2014}. The present calculations are in rather good agreement with the MC results, demonstrating the capacity of this approach to correctly predict the radial doses. Our calculations correspond to the end of the track structure, but before the shock wave is formed, and this is why they agree with the MC simulations where shock waves are not present.

Interestingly, the radial dose profile also gives the upper estimate for the pressure developed around the ion path~\cite{Surdutovich2015}.
The pressure profiles for carbon ions are also shown in Fig. \ref{fig:fig12} (the axis is given on the right). So large pressures are sufficient to initiate a shock wave in the liquid medium. The hydrodynamic equations with the initial conditions corresponding to a ``strong explosion'' were solved in Ref.~\cite{Surdutovich2010}. As it is discussed below, this solution allows one to obtain useful physical characteristics of the ion-induced shock waves which serve as a benchmark for the MD simulations~\cite{deVera2016}.

{
Besides the dose and pressure profiles, the diffusion equations also yield the initial distribution of free radicals and pre-solvated electrons~\cite{Surdutovich2015}. As a first approximation, it can be assumed that each inelastic collision leads to the formation of one OH radical. 
Under this assumption, the initial distribution of OH will follow the profile of the radial dose~\cite{Surdutovich2015}, as shown in one of the right axes in Fig. \ref{fig:fig12}. The proton transfer in the ionized water molecule dissociation is a fast process occurring by $\sim 10^{-14}$~s~\cite{Mozumder2004}. Therefore, by the time that shock waves develops ($\sim 10^{-13}$~s~\cite{Surdutovich2010,Surdutovich2015}), the radicals are almost at the same location where they were created.
}

The radial doses shown in Fig. \ref{fig:fig12} can be used to set up MD simulations of carbon ion-induced shock waves, both in and out the Bragg peak region. This is an improvement over previous simulations where the energy lost by the ion was assumed to be deposited in a cylinder of radius 1~nm~(the so-called hot cylinder)~\cite{Surdutovich2013,deVera2016}. Simulations are arranged as described in Ref.~\cite{deVera2016}, i.e., by scaling atomic velocities according to the energy deposited, but in this case using the radial dose distributions shown in Fig.~\ref{fig:fig12} for different concentric cylindrical shells around the ion path. Thus we can assess the effect of the realistic initial conditions on the shock wave development. All simulations were done using MBN Explorer~\cite{Solovyov2012}.

\begin{figure}[t]
\centering
\includegraphics[width=0.78\columnwidth]{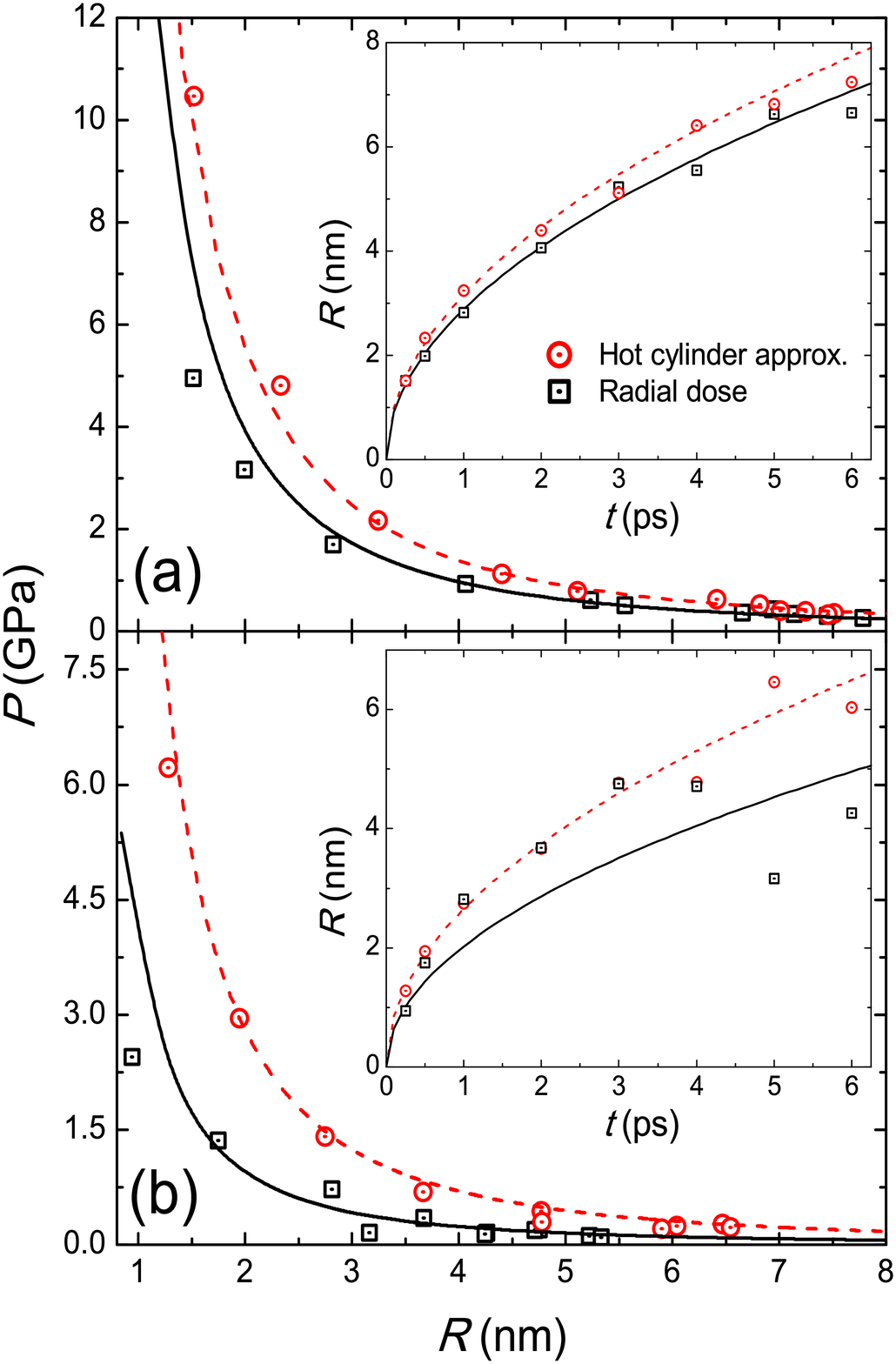}
\caption{(Color online) Pressure at the wave front as a function of its position for the shock waves produced by (a) 200-keV/u and (b) 2-MeV/u carbon ions. The insets depict the wave front position as a function of time. Symbols represent simulation results, while lines correspond to the hydrodynamic model.}
\label{fig:fig3}
\end{figure}

The dependencies of pressure at the wave front obtained from MD simulations for carbon ions in and out of the Bragg peak region on the radius of the wave front, obtained as explained in ref. \cite{deVera2016}, are shown in Fig.~\ref{fig:fig3}. The insets show the evolution of radii of wave fronts as a function of time. The results for the step-function-distributed initial pressure (hot cylinder, circles) are compared to those for the initial pressure distributed in accordance with the radial dose (squares). The hydrodynamic calculations (dashed lines)~\cite{Surdutovich2010} 
agree with the hot cylinder simulations, but the use of the radial dose 
reduces the intensity of the wave front. This reduction is up to 30 \% for a given radius in the case of the Bragg peak region (for which the hot cylinder model has been applied preciously~\cite{Surdutovich2010,Surdutovich2014,Surdutovich2015}), mainly due to the dispersion of the initial wave front, but it reaches stupendous 65 \% (for a given radius) for 2-MeV/u carbon ions, where  $\delta$-electrons become important. Out of the Bragg peak region the hydrodynamic model also reproduces the simulation results (solid lines in the figure) but assuming a reduced ``effective'' stopping power. The latter corresponds (for both energies) to the amount of energy deposited within the first $\sim 1$--$2$~nm from the ion path \cite{deVera2017}. While in the Bragg peak region most of the energy is deposited within this cylinder, this is not the case at higher energies outside the Bragg peak due to the production of more energetic $\delta$-electrons.

Once the strength of shock waves is determined, we turn our attention to their effects on the chemical stage in the Bragg peak region where they are expected to be stronger. As described above the diffusion equations also give the initial distribution of OH radicals around the ion path. This distribution can be used as the initial condition for MD simulations in which the OH transport and reaction can be included. Converting the OH concentration $n_{\rm OH}(r)$ shown in Fig.~\ref{fig:fig12} into a histogram of 5~\AA \, bin width, water molecules have been randomly selected and deprotonated, leaving a neutral OH radical. As a first approximation, only one of the most relevant reactions was considered, i.e., the OH recombination reaction:
\begin{equation}
{\rm OH} + {\rm OH} \longrightarrow {\rm H}_2{\rm O}_2 \mbox{ , }
\label{eq:OHrecomb}
\end{equation}
OH being the most important species in chemical biodamage~\cite{Douki1998}. For simplicity, in this work we disregarded other chemical reactions \cite{vonSonntag1987}, 
so pre-solvated electrons were not included in the simulations. To assure the charge neutrality, the protons coming from the water dissociation were removed from the system.

Reaction (\ref{eq:OHrecomb}) can be simulated within classical MD by virtue of the new reactive CHARMM force field introduced and implemented in MBN Explorer \cite{Sushko2016}. The CHARMM force field is very effective for simulating biomolecules in water medium~\cite{MacKerel1998} and this extension makes it possible to include chemistry merely by defining a few additional parameters, introducing reactivity almost without increasing the computational cost. The performance of this reactive force field has been shown, e.g., with the description of the water dissociation at high temperatures \cite{Sushko2016}. In the present simulations, O-H bonds, being quite stronger than O-O bonds in peroxide \cite{Ruscic2002,Blanksby2003}, are assumed to be non-reactive, while the only possible reaction considered is the formation and breaking of the O-O bond.

The parameters needed for simulation are the O-H and O-O bond distances and force constants, the O-O bond dissociation energy and cutoff distance, the atomic partial charges in the OH radical and peroxide molecule, and the equilibrium angle and force constant for the O-O-H angle \cite{Sushko2016}. Typical values have been taken from the literature: the O-H distances are similar as those in water \cite{Pabis2011} so CHARMM parameters have been kept. The values for the O-O bond and the O-O-H angle have been taken from Ref. \cite{DeGioia1999}. Partial charges of $\pm 0.375e$ have been used for OH \cite{Pabis2011} and $\pm 0.35e$ for peroxide \cite{DeGioia1999}. Inspection of the resulting O-O Morse potential \cite{Sushko2016} gives a cutoff distance of $\sim$3 \AA. The appropriateness of the parameters has been checked by comparing H$_2$O$_2$ formation G-values for 500 keV proton impact (without shock wave) to GEANT4-DNA simulations, which implement the well-known diffusion-reaction algorithms \cite{Karamitros2014}, finding good agreement at least up to $\sim 100$ ps. 

\begin{figure}[t]
\centering
\includegraphics[width=0.87\columnwidth]{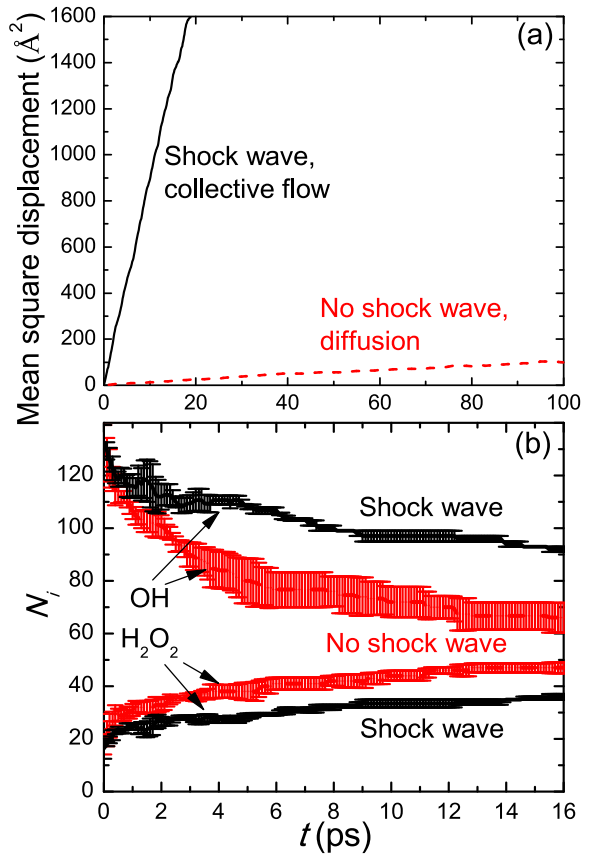}
\caption{(Color online) (a) Mean square displacement of the OH radicals produced around a 200-keV/u carbon ion path, assuming there is or there is not a shock wave. (b) Time evolution of the number of OH radicals and produced H$_2$O$_2$ molecules,
assuming there is or there is not shock wave.}
\label{fig:fig4}
\end{figure}

Figure \ref{fig:fig4} illustrates the effects of a carbon ion-induced shock wave on the OH chemistry in the Bragg peak region. Panel (a) shows the OH mean square displacements with (solid line) and without (dashed line) shock wave. In these simulations the OH recombination is not included so essentially the diffusion of OH radicals without a wave is compared to a collective flow transport with the wave.
According to the Einstein relation, the slope of the dashed line gives $6D_{\rm OH}$, so the diffusion coefficient is $D_{\rm OH}=0.166$ \AA$^2$/ps. This is in a relatively good agreement with the results of recent simulations giving 0.3 \AA$^2$/ps \cite{Pabis2011} and with the value used in the popular simulation packages PARTRAC \cite{Kreipl2009} or GEANT4-DNA \cite{Karamitros2011}, 0.28 \AA$^2$/ps. 
In the case of the shock wave the propagation is also linear with time but the slope is more than 80 times larger.
This result clearly demonstrates the ability of the shock wave to propagate free radicals (and other reactive species) more efficiently.

Finally, Fig. \ref{fig:fig4}(b) shows the results for the number of OH and H$_2$O$_2$ molecules in the simulations where the chemistry has been included, also with and without a shock wave. The lines show averages for three independent simulations with the error bars representing standard deviations. The evolutions of radicals with and without the shock wave drastically differ from each other. The shock wave prevents OH recombination, both by spreading out the radicals (as discussed above) and by creating harsh conditions in which the formation of the O-O bond is suppressed. Although only short times were simulated here, this tendency may prevail over long times. After 16 ps of transport with a wave the number of surviving OH radicals is 40\% larger than that propagated by diffusion. Even if this situation may change over longer times, it is worth noticing that the experimental G-values for OH after ns--$\mu$s-times for the largest measured LET are somewhat larger than MC simulation results, which do not include the shock waves \cite{Karamitros2014,Kreipl2009,Maeyama2011}. However, these experiments are only reported for relatively low values of LET and experiments in the Bragg peak region would be much more desirable. Such long times are still challenging for MD simulations, however G-values can be probed down to some hundreds of picosends \cite{Jonah1977}. Also, ultrafast measurements using the technique described by Dromey \textit{et al.} \cite{Dromey2016} potentially could probe the time-dependence of the OH signal on the timescales presented here, creating an opportunity to detect a chemical signature of the shock waves and directly prove their existence.

In summary, in this letter we have investigated the strength of the ion-induced shock waves inside and outside of the Bragg peak region, as well as assessed their impact on the chemistry around the ion path, by calculating realistic radial doses by means of the diffusion equations (where the contribution of $\delta$-electrons has been incorporated) and using them as the initial conditions for reactive classical MD. 
It was shown that the production of more energetic $\delta$-electrons outside of the Bragg peak region substantially weakens the shock waves, this phenomenon being confined to the Bragg peak region. The collective flow
of the shock wave in the Bragg peak region propagates the radicals 80 times faster than the diffusion mechanism, which is the only means for transport of radicals in the absence of shock waves. The waves also prevent the OH recombination to produce hydrogen peroxide. This demonstrates how the shock waves not only can produce direct physical effects (such as rupture of DNA molecules by high pressure stresses) but also modify the chemical stage of radiation damage. This fact, apart of implying a strong overlapping of the physical and chemical stages, also suggests an indirect way for experimentally detecting the ion-induced shock waves, thus far only predicted theoretically. The chemical effect has been exemplified here only by the main radiochemical reaction of OH recombination for simplicity. However, further development of the reactive force field in MBN Explorer, where other reactions can be included and/or water dissociation by the shock wave accounted for, may reveal further influence of the shock waves on the chemical stage of radiation.

The authors recognize financial support from the European Union's FP7-People Program
 within the Initial Training Network No. 608163 ``ARGENT''.







\end{document}